\documentclass[journal]{IEEEtran}
\usepackage{ulem}
\usepackage{amsmath,amssymb}

\usepackage{setspace}
\usepackage{graphicx}
\usepackage{tikz}
\usepackage{color}
\usepackage{lipsum,adjustbox}
\usetikzlibrary{decorations}
\usepackage{pgfplots,siunitx}
\usetikzlibrary{decorations.pathreplacing}
\usetikzlibrary{shapes,arrows,shadows}
\usetikzlibrary{calc,positioning}
\usepackage{tikz,tkz-euclide}

\pgfplotsset{compat=1.3}

\usetikzlibrary{calc}
\usepackage{subfig}
\usetikzlibrary{external}
\tikzsetfigurename{tikz_compiled} 

\begin{document}
\title{Comparison of Short Blocklength Sphere Shaping and Nonlinearity
Compensation in WDM Systems }

\author{Abdelkerim~Amari, Lutz Lampe, O. S. Sunish Kumar, Yunus Can G\"{u}ltekin, and Alex~Alvarado

\thanks{Abdelkerim~Amari was with Electrical and Computer Engineering Department, University of British Columbia, BC, Canada. He is now with VPIphotonics GmbH, Carnotstr. 6, 10587, Berlin, Germany. Email: abdelkerim.amari@vpiphotonics.com  } 
\thanks{Lutz Lampe and O. S. Sunish Kumar are with Electrical and Computer Engineering Department, University of British Columbia, BC, Canada.} 
\thanks{Yunus Can G\"{u}ltekin and Alex~Alvarado are with the Information and Communication Theory Lab, Signal Processing Systems Group, Department of Electrical Engineering, Eindhoven University of Technology, The Netherlands. }
}

\maketitle
\begin{abstract}
In optical communication systems, short blocklength probabilistic enumerative sphere shaping (ESS) provides both linear shaping gain and nonlinear tolerance. In this work, we investigate the performance and complexity of ESS in comparison with fiber nonlinearity compensation via digital back propagation (DBP) with different steps per span. We evaluate the impact of the shaping blocklength in terms of nonlinear tolerance and also consider the case of ESS with a Volterra-based nonlinear equalizer (VNLE), which provides lower complexity than DBP.
 In single-channel transmission, ESS with VNLE achieves similar performance in terms of finite length bit-metric decoding rate to uniform signaling with one step per span DBP. In the context of a dense wavelength-division multiplexing (WDM) transmission system, we show that ESS outperforms uniform signaling with DBP for different step sizes.
\end{abstract}
\begin{IEEEkeywords}
Digital back propagation, enumerative sphere shaping, fiber nonlinearity compensation, probabilistic shaping, optical communication systems, Volterra series.
\end{IEEEkeywords}
\IEEEpeerreviewmaketitle


\section{Introduction}
\IEEEPARstart{F}{iber} nonlinearity compensation \cite{winzer}--\cite{am1} and probabilistic shaping \cite{ps2}--\cite{KarimJLT} have been proven to be effective means to increase the spectral efficiency of optical communication systems. 
Digital back propagation (DBP) is considered as the benchmark nonlinear compensation technique due to its high performance and accuracy, when applied with small step size \cite{ip}. In wavelength-division multiplexing (WDM) transmission systems, multi-channel DBP provides the best performance and mitigates both intra-channel and inter-channel nonlinear effects \cite{mcdbp}. However, multi-channel DBP is impractical for real-time implementation due to its high complexity, its requirement for high-speed analog-to-digital converters (ADC), and also the unavailability of the information of the adjacent WDM channels. 

Probabilistic shaping based on distribution matching (DM) via constant composition (CC) \cite{ps0}, multiset partitioning \cite{mp2}, sphere shaping via shell mapping \cite{R1}, and enumerative sphere shaping (ESS) \cite{KarimJLT, KarimECOC} has been considered in optical communication systems. A combination of probabilistic shaping and DBP have been also investigated in \cite{psdbp}. In \cite{snl}, and recently in \cite{KarimJLT} and \cite{m2}, it has been shown that short blocklength shaping based on sphere shaping and CCDM provides a nonlinear tolerance gain in comparison with uniform signaling and with long blocklengths shaping. It is also known that ESS has a lower rate loss than CCDM at finite block lengths, which translates into a higher shaping gain \cite{gultekin2,GISIT}. This makes short blocklength ESS an interesting approach for achieving both shaping gain and nonlinear tolerance.

In this paper, we propose to exploit the nonlinear tolerance gain that short block length ESS provides \cite{KarimJLT, KarimECOC}, as a way to avoid the use of high complexity nonlinearity compensation techniques like DBP. We investigate the performance of uniform signaling with fiber nonlinearity compensation techniques and ESS with and without nonlinearity compensation, and explore the possibility of complexity reduction in this context. 
We compare the performance of ESS with different blocklengths to uniform signaling with nonlinearity compensation via single-channel DBP. 
We also consider the case of ESS with Volterra based nonlinear equalization (VNLE) \cite{VNLE1, VNLE2}, which reduces the complexity of nonlinear compensation by half, when compared to DBP \cite{VNLE1}. An evaluation of the complexity and storage of the proposed shaping and nonlinearity compensation approaches is also performed.

For a dense WDM system, we show that ESS provides better performance in terms of finite length bit-metric decoding (BMD) rate when compared to uniform with DBP for different number of steps per span. In terms of nonlinear tolerance, ESS exhibits the highest effective signal-to-noise ratio (SNR) at the shortest block length. In this context, ESS with VNLE and ESS with DBP applied at one step per span exceeds the performance of uniform with DBP per span and with $4$ steps per span DBP, respectively.
ESS also has lower complexity than DBP. However, ESS introduces additional latency and storage requirements. These results identify the potential for complexity reduction when considering short blocklength ESS instead of fiber nonlinearity compensation via DBP. 
\section{System model and performance metrics}\label{sec:SP}
\subsection{System Model}\label{sec:SPmodel}
\begin{figure*}[tbp]
	\centering		
		\includegraphics[width=0.75\linewidth]{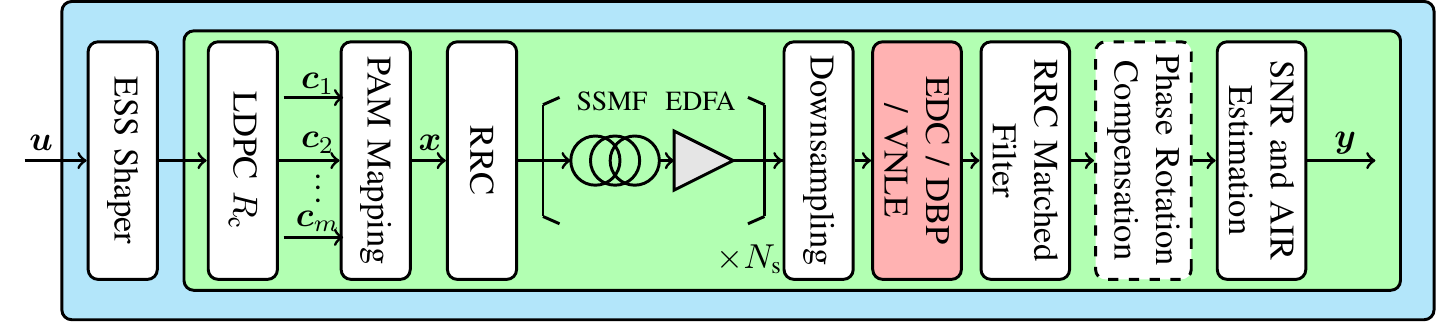}
		\caption{Transmission diagram.
		$N_s$: number of spans. Dashed box is used for only CD compensation. Green color and cyan color correspond to uniform and shaping signaling respectively.
	}
		\label{fig:2}
		\vspace{-0.3cm}
\end{figure*}

The block diagram of the considered system is shown in  Fig.~\ref{fig:2}. The information bits are shaped based on ESS and passed through a rate $R_\text{c}$ low-density parity-check (LDPC) encoder following the probabilistic amplitude shaping (PAS) framework \cite{pas}. 
The ESS shaper generates bounded energy sequences by fixing a maximum energy constraint \cite{gultekin2, KarimJLT}. The shaping rate $R_\text{s}=k/N$~[bits/amp], where $k$ and $N$ are the number of input information bits and output amplitudes respectively, is obtained by adjusting the maximum energy constraint. 
 The shaping rate used in this work is $R_\text{s}=1.85$ [bits/amp.]. The LDPC parity bits and a part of the information bits are used as sign bits for the shaped amplitudes, as explained in [4]. 64-quadrature amplitude modulation (QAM) (8-Pulse-amplitude modulation (PAM) for shaping signaling with amplitude alphabet cardinality $a=4$, applied per real dimension) is used as modulation format for both shaped and uniform signaling due to its high spectral efficiency and the considerable possible shaping gain, which increases with the modulation format. Different LDPC rates $R_\text{c} = 3/4$ and $R_\text{c}=4/5$ should be used for uniform and shaped signaling, respectively, to ensure a fair comparison with the same information rate $R =2.25$ [bits/1D-sym]. In this work we focus on the SNR and AIR performances.

At the receiver side, nonlinear compensation is applied via DBP or VNLE. We also consider the case of only linear chromatic dispersion (CD) compensation. 

\vspace{-0.2cm}
\subsection{Performance Metrics}\label{sec:SPerf}
 It has been shown that short blocklengths shaping provides interesting performance in terms of nonlinear tolerance in the optical fiber channel \cite{KarimJLT, KarimECOC, m2}. Thus, suitable performance metrics like finite length BMD rate, taking into account the rate loss with respect to infinite blocklengths, and effective SNR are used in this work. The finite length BMD rate gives an indication of the overall performance including the nonlinear gain and the linear shaping gain, and the effective SNR measures only the nonlinear tolerance gain.
The finite length BMD rate is defined as \cite{ps2}
 \begin{equation}\label{gmi}
     \text{AIR}_{\textit{N}} = \underbrace{\left[  H(\boldsymbol{C}) - \sum_{i=1}^m H(C_i \mid Y) \right]}_{\text{BMD Rate}} - \underbrace{\left[H(A)-\frac{k}{N}\right]}_{\text{Rate loss}},
 \end{equation}
where $H(\cdot)$ denotes entropy, $m$ is the number of bits per symbol, and \textit{$A$} are the shaped amplitudes. $\boldsymbol {C}= ( C_1, C_2, ..., C_m )$ are the bit levels of the transmitted symbol, and $Y$ corresponds to the received symbol.
The rate loss corresponds to the gap between the entropy and the shaping rate, and vanishes at infinite blocklengths. 

The effective SNR includes both amplified spontaneous emission (ASE) noise and nonlinear noise contributions. It is calculated per QAM symbol taking into account the probability of each constellation point, and defined as \cite{ps3} 
\begin{align}\label{eff.snr}
\text{SNR}_{\text{eff}} = \frac{\mathbb{E}[|{X}|^2]}{\mathbb{E}[|{Y}-{X}|^2]}, 
\end{align}
where $\mathbb{E}[\cdot]$ represents expectation, and $X$ and $Y$ are the transmitted and received symbols respectively (see Fig.~\ref{fig:2}). 

\section{Simulation setup and results}\label{sec:simus}
\subsection{Simulation Setup}\label{sec:simusset}
 We consider two simulation scenarios: dual-polarization single-channel system, where the simulation setup is shown in Fig.~\ref{fig:2}, and dual-polarization 11 WDM channels system to quantify the effect of intra-channel and inter-channel nonlinear effects, respectively.
We compare the performance of ESS with linear electronic dispersion compensation (EDC) and ESS with nonlinearity compensation via VNLE and DBP against uniform signaling with DBP applied with different number of steps per span. The VNLE is performed in parallel and applied once per span \cite{VNLE1}.
In this work, polarization mode dispersion and linear phase noise are neglected to focus on the impact of the nonlinear effects. When EDC only is performed, we assume an ideal compensation of the common phase rotation of the entire constellation due to nonlinearity.
We consider a dense WDM scenario with large symbol rate and high order modulation format to ensure a transmission with high data rate. The symbol rate is $45$ Gbaud. The modulation format is $64$ QAM. We use a root-raised cosine (RRC) filter with a roll-off factor $\rho = 0.1$.

We consider a dispersion unmanaged system with multi-span standard single-mode fiber (SSMF). Concerning the SSMF parameters, the attenuation coefficient is $\alpha=0.2~\mathrm{dB\cdot km^{-1}}$, the dispersion parameter is $D=17 ~\mathrm{ps \cdot nm^{-1} \cdot km^{-1}}$, and the nonlinear coefficient is \mbox{$\gamma=1.3~ \mathrm{W^{-1} \cdot km^{-1}}$}. The signal is amplified after each $L=80$~km span by an erbium-doped fiber amplifier (EDFA) with a $5$ dB noise figure and $16$ dB gain.
At the receiver side, the signal is passed by a channel selection, and the nonlinear compensation is applied after downsampling to $2$ samples/symbol.

\subsection{Simulation Results}\label{sec:simusres}
\begin{figure}[!t]
		\includegraphics[width=0.83\linewidth]{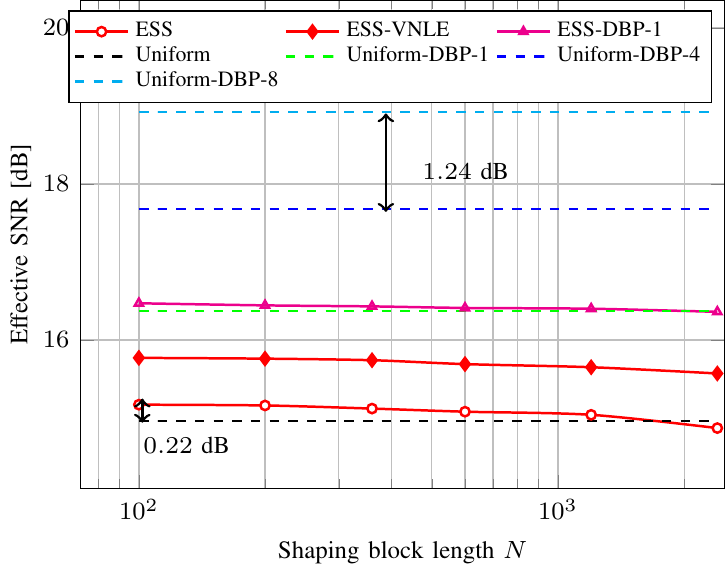}
		\caption{Effective SNR vs. shaping blocklength $N$ for a single WDM channel.}	
		\label{fig:3}
				\vspace{-0.2cm}
\end{figure}

We firstly consider the dual-polarization single-channel system. In Fig.~\ref{fig:3}, we plot the effective SNR versus shaping blocklength $N$ for a transmission reach of $2800$ km at optimal input power. Fig.~\ref{fig:3} shows that uniform signaling with DBP applied at $8$ steps per span exhibits the best performance and the gain is $1.24$ dB, $2.55$ dB, and $3.96$ dB in comparison with DBP at $4$ steps per span, DBP applied per span and EDC-only, respectively. ESS exhibits its best performance at the shortest blocklength, and for $N=100$, it shows a gain of $0.22$~dB in comparison with uniform signaling. At the same blocklength, ESS with one step per span DBP exhibits a gain of $0.1$~dB in comparison with uniform with DBP applied per span.

\begin{figure}[!]
	\includegraphics[width=0.83\linewidth]{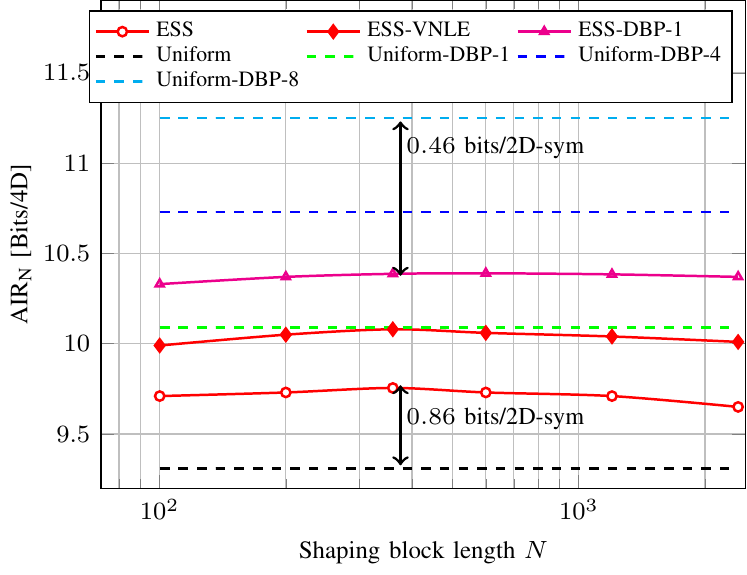}
		\caption{Finite length AIR vs. shaping blocklength $N$ for a single WDM channel.}	
		\label{fig:4}
		\vspace{-0.2cm}
\end{figure}

Fig.~\ref{fig:4} shows the finite length BMD rate performances as a function of the shaping blocklength.
The highest performance of ESS with VNLE is obtained at $N=360$. It corresponds to the optimal trade-off between the nonlinear tolerance, which is inversely proportional to the blocklength, and the linear shaping gain, which increases with the blocklength. For this blocklength, ESS with nonlinearity compensation via VNLE provides similar performance to uniform with DBP applied per span, which allows a reduction of nonlinear compensation complexity by half \cite{am1}. DBP at $8$ steps per span exhibits the best performance and shows a gain of $0.52$~dB in comparison with DBP at $4$ steps per span, due to its higher accuracy.

Next, we extend the scenario to a $11$ dual-polarization WDM channels system with $45$ Gb in $50$ GHz grid configuration. The results for the middle WDM channel will be shown because it is the most affected by the nonlinear impairments.

\begin{figure}[!t]
		\includegraphics[width=0.83\linewidth]{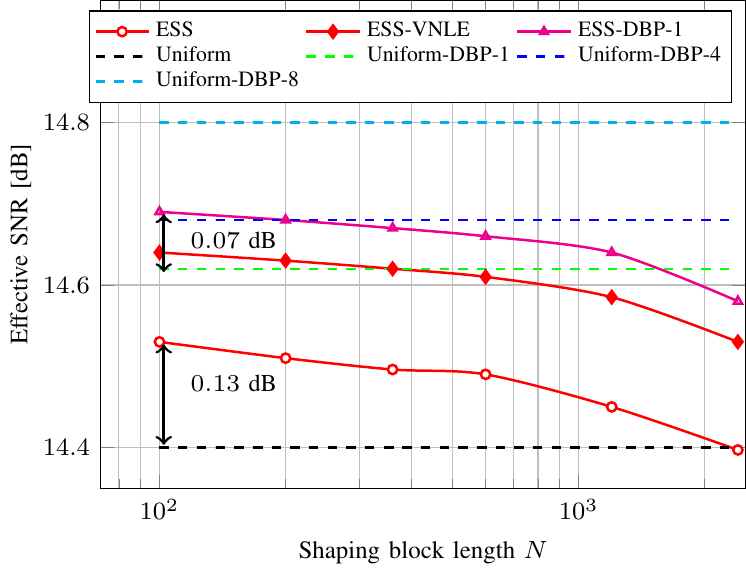}
		\caption{Effective SNR vs. shaping blocklength $N$ for $11$ WDM channels.}	
		\label{fig:5}
						\vspace{-0.2cm}
\end{figure}

In Fig.~\ref{fig:5}, the effective SNR is plotted as a function of the shaping block length for $2000$ km as transmission distance at optimal input power. It is observed that $64$-QAM uniform signaling with DBP at $8$ steps per span provides the best performance, and the gain is about $0.4$ dB, $0.18$ dB and $0.12$ dB in comparison with CD compensation, DBP applied per span and DBP at $4$ steps per span, respectively. Again, it is shown that ESS gives the best performance at the shortest blocklength, while at a blocklength $N=2400$, uniform signaling shows better performance than ESS. It is also observed that the perfromance gap between ESS and ESS with DBP applied per span is lower than the case of uniform and uniform with DBP applied per span. This can be explained by the fact that ESS with short bloclengths mitigates a part of the nonlinearity and also has different statistics than uniform signaling, which results on different behaviors of DBP for both signaling.
For $N=100$, ESS with DBP and ESS with VNLE, applied per span, provide gains of about $0.02$ dB and $0.01$~dB when compared to uniform signaling with DBP at $4$ per span and DBP per span, respectively. This means that in the presence of probabilistic shaping, by using short blocklength shaping, the complexity of the nonlinearity compensation can be significantly reduced in dense WDM transmission systems.

\begin{table*}[t]
\begin{centering}
\caption{Computational and storage complexity.}
\par\end{centering}
\begin{centering}
{\scriptsize{}}%
\begin{tabular}{|>{\centering}m{0.12\paperwidth}|>{\centering}m{0.2\paperwidth}|>{\centering}m{0.3\paperwidth}|>{\centering}m{0.1\paperwidth}|}
\cline{2-4} \cline{3-4} \cline{4-4} 
\multicolumn{1}{>{\centering}m{0.12\paperwidth}|}{} & {\scriptsize{}No. of real-valued multiplications} & {\scriptsize{}No. of real-valued additions} & {\scriptsize{}Storage requirement}\tabularnewline
\hline 
\centering{}{\scriptsize{}Uniform} & {\scriptsize{}$(8N_\text{FFT}\log_{2}\left(N_\text{FFT}\right)+8N_\text{FFT})/N_\text{sym}$} & {\scriptsize{}$(8N_\text{FFT}\log_{2}\left(N_\text{FFT}\right))/N_\text{sym}$} & {\scriptsize{}-}\tabularnewline
\hline 

\centering{}{\scriptsize{}Uniform-DBP} & {\scriptsize{}$(8N_\text{s}sN_\text{FFT}\log_{2}\left(N_\text{FFT}\right)+21N_\text{s}sN_\text{FFT})/N_\text{sym}$} & {\scriptsize{}$(8N_\text{s}sN_\text{FFT}\log_{2}\left(N_\text{FFT}\right)+4N_\text{s}sN_\text{FFT})/N_\text{sym}$} & {\scriptsize{}-}\tabularnewline
\hline 
\centering{}{\scriptsize{}ESS} & {\scriptsize{}$(8N_\text{FFT}\log_{2}\left(N_\text{FFT}\right)+8N_\text{FFT})/N_\text{sym}$} & {\scriptsize{}$(8aN_\text{sym}$+$8N_\text{FFT}\log_{2}\left(N_\text{FFT}\right))/N_\text{sym}$} & {\scriptsize{} $O(N^{2}\,\log\,N)$}\tabularnewline
\hline 
\centering{}{\scriptsize{}ESS-DBP} & {\scriptsize{}$(8N_\text{s}sN_\text{FFT}\log_{2}\left(N_\text{FFT}\right)+21N_\text{s}sN_\text{FFT})/N_\text{sym}$} & {\scriptsize{}$8aN_\text{sym}$+$8N_\text{s}sN_\text{FFT}\log_{2}\left(N_\text{FFT}\right)+4N_\text{s}sN_\text{FFT})/N_\text{sym}$} & {\scriptsize{}$O(N^{2}\,\log\,N)$}\tabularnewline
\hline 
\centering{}{\scriptsize{}ESS-VNLE} & {\scriptsize{}$(8N_\text{s}N_\text{FFT}\log_{2}\left(N_\text{FFT}\right)+8.5N_\text{s}N_\text{FFT}+4N_\text{FFT}+4N_\text{FFT}\log_{2}\left(N_\text{FFT}\right))/N_\text{sym}$} & {\scriptsize{}$(8aN_\text{sym}$+$8N_\text{s}N_\text{FFT}\log_{2}\left(N_\text{FFT}\right)+8N_\text{s}N_\text{FFT}+8N_\text{FFT}\log_{2}\left(N_\text{FFT}\right)+4N_\text{FFT})/N_\text{sym}$} & {\scriptsize{} $O(N^{2}\,\log\,N)$}\tabularnewline
\hline 

\end{tabular}{\scriptsize\par}
\par\end{centering}
\vspace{1mm}

\centering{}{\scriptsize{}$N_\text{s}:$ Number of fiber spans, $s:$ Number
of steps per fiber span, $N_\text{FFT}:$ Fast Fourier transform size, $a$: amplitude alphabet cardinlaity, $N_\text{sym}:$ Total number of symbols.}{\scriptsize\par}
\end{table*}

\begin{figure}
			\includegraphics[width=0.83\linewidth]{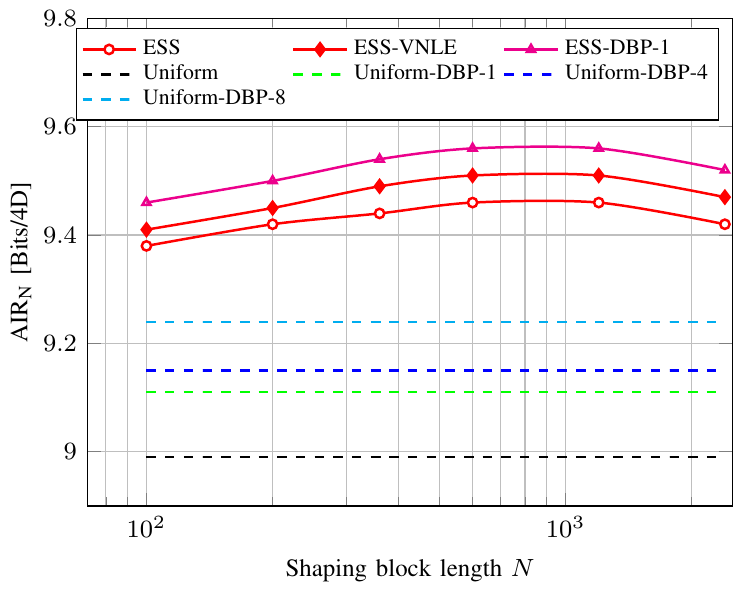}
	\caption{Finite length AIR vs. shaping blocklength $N$ for $11$ WDM channel.}	
	\label{fig:6}
\end{figure}

In terms of the finite length BMD rate, as shown in Fig.~\ref{fig:6}, ESS with DBP applied per span exhibits the best performance. The shaping blocklength that provides the optimal trade-off between linear shaping gain and nonlinear tolerance is around $N=600$. It is also observed that DBP with a high number of steps per span, i.e., high accuracy, still shows lower performance than ESS with only linear CD compensation.
\vspace{-0.2cm}
\subsection{Complexity Analysis}

Table I summarizes the computational complexity and the required storage for the considered techniques. The computational complexity is evaluated for a 4-dimensional symbol (i.e., dual-polarization symbols). It is important to mention that the uniform signaling and ESS only cases require the CD compensation, and the significant portion of the complexity for such techniques comes from the CD compensation part. On the other hand, when these techniques (i.e., uniform signaling and ESS) are combined with DBP or VNLE, the CD compensation part is already included in the DBP and VNLE implementation. Bounded-precision ESS \cite{GISIT} is used in this work to reduce the storage requirements \cite{cmpx}. The CD compensation is implemented in frequency-domain using a fast Fourier transform (FFT)/Inverse-FFT method, as in \cite{VNLE1}. For both DBP and VNLE methods, we follow the same frequency-domain approach for CD compensation. 
The ESS can be implemented with a smaller computational complexity than nonlinearity compensation due to its lower number of real-valued multiplications and additions. In addition, ESS complexity does not depend on the number of spans, unlike nonlinear compensation via DBP and VNLE.  However, its realization requires additional storage. There is a trade-off between the computational complexity and the required storage for the shaping and nonlinearity compensation techniques. With short blocklength ESS, the nonlinearity compensation complexity can be significantly reduced with increased performance, especially for the  WDM systems.
\section{Conclusion}\label{sec:con}
 We have investigated the performance of fiber nonlinearity compensation in comparison with finite blocklength ESS in single-channel and dense WDM transmission systems. We have shown that ESS exceeds the performance of uniform signaling with higher complexity nonlinearity compensation in terms of  finite length bit-metric decoding rate. Furthermore, shorth blocklengths ESS, which provide nonlinear tolerance gain, has lower complexity in terms of real-valued multiplications and additions than DBP, but it introduces storage requirements. This make short blocklength probabilistic shaping more suitable for high data rate dense WDM systems than nonlinear effects compensation via DBP.
 

\end{document}